\documentclass[journal]{IEEEtran}
\usepackage{tikz}
\usepackage{pgfplots}
\usepackage{hyperref}
\usepackage{graphicx}
\usepackage{bm,bbm}
\usepackage{array}
\usepackage{color, colortbl}
\definecolor{Gray}{gray}{0.95}
\usepackage{siunitx}
\usepackage{lineno}
\usepackage{multirow}
\usepackage{cite}
\usepackage[caption=false, font=footnotesize, margin=.5cm]{subfig}
\usepackage{cuted}
\usepackage[normalem]{ulem}
\usepackage{float}
\usepackage{dblfloatfix}
\usepackage[export]{adjustbox}
\usepackage{dblfloatfix}
\usepackage{booktabs,tabularx}
\usepackage{scalerel}

\usetikzlibrary{svg.path}

\definecolor{orcidlogocol}{HTML}{A6CE39}
\tikzset{
	orcidlogo/.pic={
		\fill[orcidlogocol] svg{M256,128c0,70.7-57.3,128-128,128C57.3,256,0,198.7,0,128C0,57.3,57.3,0,128,0C198.7,0,256,57.3,256,128z};
		\fill[white] svg{M86.3,186.2H70.9V79.1h15.4v48.4V186.2z}
		svg{M108.9,79.1h41.6c39.6,0,57,28.3,57,53.6c0,27.5-21.5,53.6-56.8,53.6h-41.8V79.1z M124.3,172.4h24.5c34.9,0,42.9-26.5,42.9-39.7c0-21.5-13.7-39.7-43.7-39.7h-23.7V172.4z}
		svg{M88.7,56.8c0,5.5-4.5,10.1-10.1,10.1c-5.6,0-10.1-4.6-10.1-10.1c0-5.6,4.5-10.1,10.1-10.1C84.2,46.7,88.7,51.3,88.7,56.8z};
	}
}

\newcommand\orcidicon[1]{\href{https://orcid.org/#1}{\mbox{\scalerel*{
				\begin{tikzpicture}[yscale=-1,transform shape]
				\pic{orcidlogo};
				\end{tikzpicture}
			}{|}}}}

\ifCLASSINFOpdf
\else
\fi
\usepackage[cmex10]{amsmath}
\begin{document}
%


\title{Deriving Surface Resistivity from Polarimetric SAR Data Using Dual-Input UNet}
%
%



\author{Bibin Wilson\textsuperscript{*},
Rajiv Kumar\textsuperscript{!},
Narayanarao~Bhogapurapu\textsuperscript{\#},
Anand Singh\textsuperscript{*}, Amit Sethi\textsuperscript{\textdagger}
\thanks{{\textsuperscript{*}Department of Earth Sciences,
\textsuperscript{!}Department of Computer Science and Engineering,
\textsuperscript{\#}Microwave Remote Sensing Lab, Center of Studies in Resources
Engineering,
\textsuperscript{\textdagger}}Department of Electrical Engineering,
\\Indian Institute of Technology Bombay, Maharashtra, India}
}

\maketitle

\begin{abstract}
Traditional
survey methods for finding surface resistivity are time consuming and labor intensive. Very few studies have focused
on finding the resistivity / conductivity using remote sensing data
and deep learning techniques. In this line of work, we assessed the correlation between surface resistivity and Synthetic Aperture Radar (SAR) by applying various deep learning methods and tested our hypothesis in the Coso Geothermal Area, USA.
For detecting the resistivity, L-band full polarimetric SAR data acquired by UAVSAR were used, and MT (Magnetotellurics) inverted resistivity data of the area were used as the ground truth. We conducted experiments to compare various deep learning architectures, and suggest the use of Dual Input UNet (DI-UNet) architecture. DI-UNet uses a deep learning architecture to predict the resistivity using full polarimetric SAR data by promising a quick survey addition to the traditional method. Our proposed approach accomplished improved outcomes for the mapping of MT resistivity from SAR data.
\end{abstract}


%
\IEEEpeerreviewmaketitle

\section{Introduction} \label{sec:intro}


\IEEEPARstart{M}{any} 
satellites incorporate suborbital multi-spectral, hyper-spectral, and SAR sensors with global coverages. This provides short-time threshold investigations and repeated-pass data at much shorter time frames~\cite{landuyt2018flood}. Due to this factor, there are multiple studies that derive field measurements from remote sensing data instead of manual field data collection. This includes air quality measurements using satellite-derived particulate matter measurements~\cite{gui2019satellite}, rainfall observation~\cite{kimani2017assessment}, bathymetry~\cite{wilson2020satellite}, climatology~\cite{shikhov2018satellite}, chlorophyll and algae bloom indices~\cite{binding2018analysis}, land surface temperature~\cite{deo2017forecasting}, etc.

Synthetic Aperture Radar (SAR) is an active microwave remote sensing technology that uses the difference in delays of electromagnetic waves that are emitted and then received to generate images. They are typically mounted on aircrafts or space-based satellites. Due to its independence in air and sunlight circumstances, SAR frames are of considerable scientific importance ~\cite{gao2019transferred}. The growing availability of SAR data leads to multiple scientific studies including flood mapping ~\cite{landuyt2018flood,clement2018multi}, land mine detection ~\cite{fernandez2018synthetic}, canola phenology estimation ~\cite{mcnairn2018estimating}, landslide detection ~\cite{mondini2021landslide}, target recognition ~\cite{ding2017target}, flood inundation mapping ~\cite{shen2019near}.

In recent years, deep learning~\cite{DL_survey} has been explored rapidly in several disciplines~\cite{DL_scientific_disc} such as computer vision ~\cite{rajiv_siva}, speech recognition and natural language processing~\cite{transformers_survey}, ~\cite{DL_in_RS_survey}. In contrast to traditional machine learning algorithms ~\cite{MLvsDL}, deep learning based approaches often use multiple processing layers to learn data representations with multiple levels of abstraction~\cite{lecun2015deep}. A convolutional neural network (CNN) can discover features from images with multiple layers and accumulate the low to high-level characteristics. They can then be applied to different computer vision activities, such as large-scale picture identification, object detection and sensing,  semantic segmentation~\cite{voulodimos2018deep}.

Subsurface resistivity surveys are a standard method for geophysical subsurface mapping and are broadly applied to detect minerals, hydrology, archaeological planning, atmospheric and civil infrastructure, etc. This approach is one of the most popular geophysical methods to identify land lift characteristics, such as the thickness of sliding bodies, lateral expansion, thickness, and spatially and temporally available water content modification~\cite{burger1992exploration}. Magnetotelluric (MT) resistivity survey method is a passive electromagnetic technique that measures the magnetic and electric field on the earth's surface in an orthogonal direction. MT studies were used in wide range of studies, such as to identify shallow and deep aquifers ~\cite{yadav2020application}, to image carbon dioxide storage sites, gas shales ~\cite{streich2010imaging}, geothermal reservoirs, etc. 

The main drawback of traditional resistivity survey method is that it is time consuming, labor intensive, and require costly equipment. Although we need to have a traditional resistivity survey for subsurface studies, we believe that an approach to derive the surface resistivity from remote sensing data will give a jump-start to traditional approaches by providing prior information on the surface resistivity. 

Multiple images of the same location separated in the frequency or time domain~\cite{tebaldini2010methods} from each orbit enables multiple degrees of freedom for conducting experiments and avoid dependency on a single incidence angle.
Although there are studies on the prediction of soil electrical conductivity using remote sensing ~\cite{rahmati2017quantitative}, ~\cite{shrestha2006relating}, very few studies have focused on finding resistivity / conductivity using remote sensing and deep learning techniques. In this paper, we derive the surface resistivity using Synthetic Aperture Radar data. Due to the limitations of penetration capabilities of radar data, we restrict the scope of our study to surface resistivity prediction, though resistivity surveys such as DC (Direct Current), MT are available for both surface and subsurface analysis.


The contributions of this paper are three-fold:
\begin{enumerate}
\item We propose a Dual-Input UNet architecture to estimate surface resistivity from SAR data.
\item We demonstrate the application of Dual-Input UNet for surface resistivity retrieval from  multi-incidence angle airborne full-polarimetric L-band SAR data.

\item We also show that segmentation-based architectures are ideal for solving the problem, by implementing Dual-Input UNet and comparing the results with U-Net~\cite{ronneberger2015u}, SegNet~\cite{badrinarayanan2017segnet}, and PSPNet~\cite{zhao2017pyramid} architectures.


\end{enumerate}

The rest of the article is structured as follows: Section~\ref{sec:methodology} covers our proposed methodology by introducing SAR and MT literature and the proposed model architecture in a step by step manner. Section~\ref{sec:datasetandimplementation} consists of subsection~\ref{subsec:dataset} that explains the details of the dataset and subsection~\ref{subsec:Implementation details} that gives the implementation details. Section \ref{sec:experimentsandevaluation} discusses about the various evaluation metrics and error metrics followed by the experiments subsection. Section~\ref{sec:results} discusses about the results, its discussion and analysis. Finally, the conclusion, followed by the future scope are briefly mentioned in Section~\ref{sec:conclusion}.
\color{black}

\section{Methodology} \label{sec:methodology}

The main objective of this research is to estimate surface resistivity from remote sensing SAR data, by leveraging the multi-incidence angle SAR data using deep learning.

\begin{figure*}[t!] \label{fig:unet}
\centering     
\includegraphics[width=0.9\textwidth]{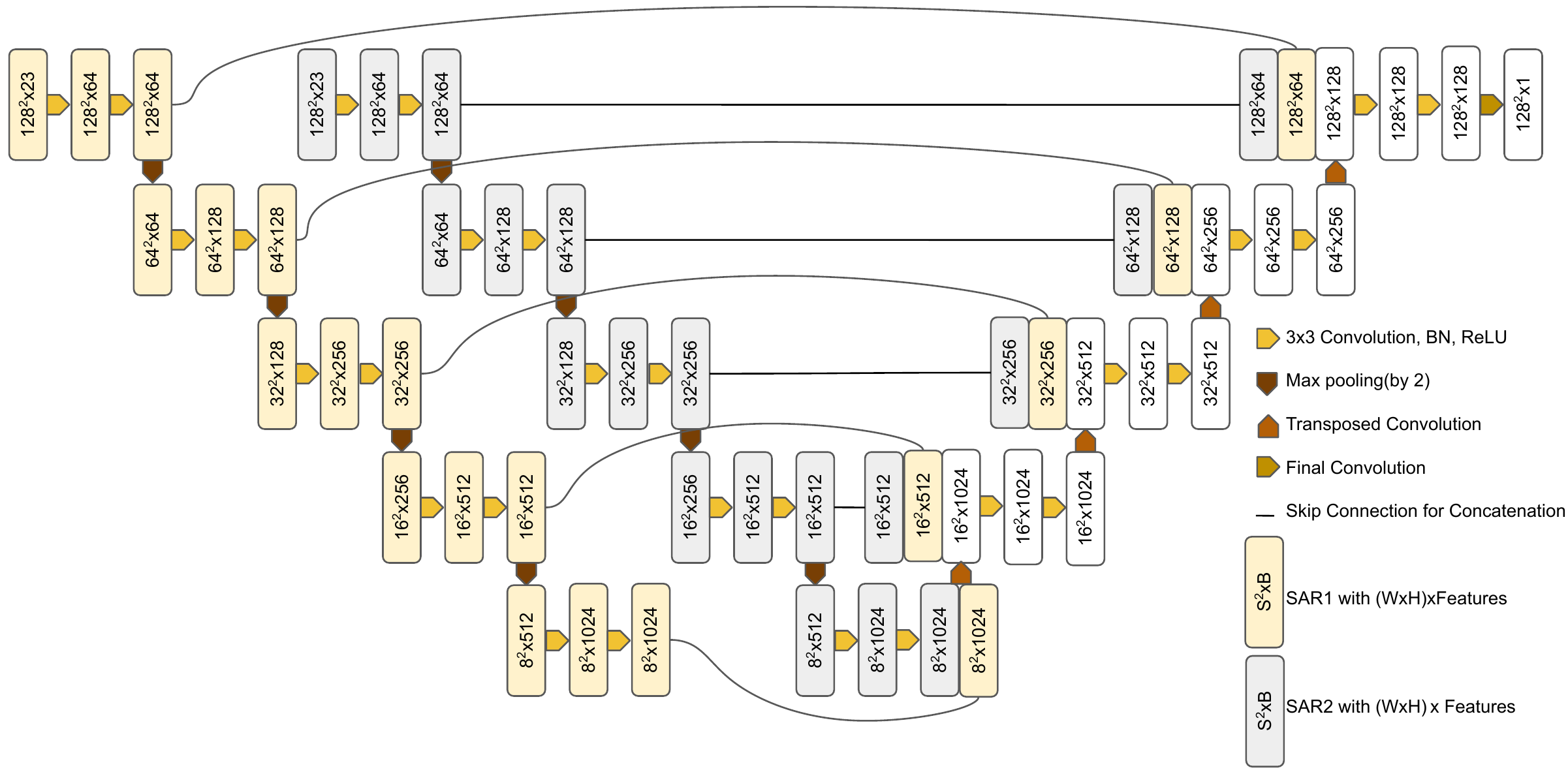}
\caption{DI-UNet: Network Architecture of our Method, having 2 encoder paths and one decoder. Input to the encoders are the polarimetric features derived from the UAVSAR. Encoders concatenated with the decoder through skip connections.Each block having dimension $S^2$ x B where S is the spatial dimension(HxW) and B is the number of bands(features). Detailed implementation details in section\ref{ref:implementation}}
\label{fig:DIUNet}
\end{figure*}

\subsection{Relation between Resistivity/Conductivity with SAR data}



\subsubsection{Plane Electro Magnetic Waves}

As part of the mathematical analysis of electromagnetic (EM) wave propagation and reflection, it is common to incorporate three fundamental properties of the medium: dielectric permittivity, $\varepsilon$; electric conductivity, $\sigma$; and magnetic permeability, $\mu$. These wave qualities are illustrated by Maxwell's equations for EM waves in homogeneous and isotropic mediums~\cite{stratton1941electromagnetic} given by, 



\noindent\begin{minipage}{.5\linewidth}
\begin{align}
    \bm{\nabla}\cdot\bm{E} &=0\;,\label{eq:Maxwell1}
    \vphantom{\frac{\partial\bm{B}}{\partial t}}\\
    \bm{\nabla}\cdot\bm{H} &=0\;,\vphantom{\frac{\partial\bm{B}}{\partial t}}\label{eq:Maxwell2}
\end{align}
\end{minipage}%
\begin{minipage}{.5\linewidth}
\begin{align}
    \bm{\nabla}\times\bm{E} &=-\frac{\partial\bm{B}}{\partial t}\;,\label{eq:Maxwell3}
    \\
    \bm{\nabla}\times\bm{H} &=\bm{J}_f
    +\frac{\partial\bm{D}}{\partial t}\;,\label{eq:Maxwell4}
\end{align}
\end{minipage}
\smallskip


\noindent
where $E$ denotes the vector for electric field intensity, $H$ denotes the magnetic field intensity vector, and t denotes the time.


Both magnetic permeability and dielectric permittivity are functions of time derivatives of the magnetic and electric field intensities, respectively. Even in free space, their values are never zero. When discussing electrical conductivity in the geophysical sciences, it is a common practice to refer to it in terms of electrical resistance. This means that the permeability of the Earth's surface and near-surface materials such as water and common minerals is not appreciably different from that of free space. This will allow us to remove magnetic permeability from radar wave reflection calculations, in which it occurs as both a numerator and denominator. Using only two material qualities, dielectric permittivity and electrical conductivity/resistivity, the complex problem of radar reflection will be simplified.~\cite{o1976magnetic}


Harmonic plane waves in homogeneous and isotropic mediums can be propagated and reflected using a complex propagation constant, $k$, defined as

\begin{equation} \label{eq22}
    k=\alpha+i \beta,
\end{equation}

\noindent
where $\alpha$ is the phase constant and $\beta$ is the attenuation factor, while $i$ is the standard imaginary unit, such that $i^2={-}1$ and can be expressed as given in Eq.~\eqref{eq6} and~\eqref{eq7}.

\begin{equation} \label{eq6}
    \alpha = \omega \left[\frac{\mu \varepsilon}{2} \left( \sqrt{1 + \frac{\sigma^2}{\varepsilon^2 \omega^2}    } +1 \right) \right]^{1/2},
\end{equation}
and 
\begin{equation} \label{eq7}
    \beta = \omega \left[\frac{\mu \varepsilon}{2} \left( \sqrt{1 + \frac{\sigma^2}{\varepsilon^2 \omega^2}    } -1 \right) \right]^{1/2},
\end{equation}
$\omega$ is the angular frequency, and $\omega = 2 \pi f$ where $f$ is the linear frequency,

\subsubsection{General and simplified forms of the radar reflection coefficient}

Three scalar material qualities separate two homogeneous and isotropic halves spaces at a reflecting interface: In this case, $ \varepsilon_1, \varepsilon_2, \sigma_1, \sigma_2, \mu_1, \mu_2$. A plane wave entering the interface at normal incidence from medium 1 toward medium 2 is solely considered for specular reflection~\cite{stratton1941electromagnetic}.

\begin{equation} \label{eq7a}
    r = \frac{E_r}{E_o} = \frac{\mu_2 k_1 - \mu_1 k_2}{\mu_2 k_1 + \mu_1 k_2},
\end{equation}
\noindent

Complex reflection coefficient($r$), defined as the complex intensity of reflected wave $E_r$, normalized by the complex intensity of the incident wave, $E_o$. Equations Eqns.\eqref{eq22}, \eqref{eq6}, and \eqref{eq7} relate the complex propagation constants $k_1$ and the $k_2$ on both sides of the reflecting interface to their material constants.


We shall assume that both media have the magnetic permeability of free space from here on out, as it is fair to do for most rocks and minerals at temperatures and pressures around the Earth's surface. As a result of this reduction, Eq.~\eqref{eq7a} can be written as

\begin{equation}\label{eq:solv1}
    r = \frac{k_1 - k_2}{k_1 + k_2} = \frac{\alpha_1+i\beta_1-\alpha_2-i\beta_2}{\alpha_1+i\beta_1+\alpha_2-i\beta_2}.
\end{equation}

Here we have extended our equation to include the propagation constant of each medium, $k_1$ and $k_2$. The real amplitude reflection coefficient, can be expressed simply as the absolute value of the complex vector $r$:
\begin{equation} \label{eq:solv2}
    r = \left|r\right| = \sqrt{\frac{(\alpha_1-\alpha2)^2+(\beta_1-\beta_2)^2}{(\alpha_1+\alpha2)^2+(\beta_1+\beta_2)^2}}.
\end{equation}
Pythagorean length of complex vector, $r$, is defined to be the absolute value of its length in the complex plane (Argand diagram). $\phi$, the tangent of the phase shift angle of the reflected wave is determined by:
\begin{equation}
    tan(\phi) = \frac{2(\alpha_2\beta_2 - \alpha_1\beta_2)}{(\alpha_1^2+\beta_1^2)-(\alpha_2^2+\beta_2^2)}.
\end{equation}

Since it is possible to estimate the radar's reflectivity and phase shift with sufficient precision, Eq.\eqref{eq:solv1} and Eq.\eqref{eq:solv2} can be viewed as a pair of equations with two unknowns, $\sigma_2$ and $\varepsilon_2$, respectively.


Next, we'll look at Eq.\eqref{eq:solv1}, which has low loss and high loss as its two limiting cases. There are two ways to look at this: in the first case, $\sigma_2 << \varepsilon_2\omega$ we get:

\begin{equation}
    r = \sqrt{\frac{(\alpha_1-\alpha_2)^2}{(\alpha_1+\alpha_2)^2}} = \frac{\alpha_1-\alpha_2}{\alpha_1+\alpha_2} = \frac{\sqrt{\varepsilon_1}-\sqrt{\varepsilon_2}}{\sqrt{\varepsilon_1}+\sqrt{\varepsilon_2}}.
\end{equation}


Air permittivity ($\varepsilon_1$) and surface material ($\varepsilon_2$) are all that is needed to calculate reflection coefficient.


We suppose that the air (medium 1) is still a lossless dielectric, but that the surface medium (medium 2) is a high loss, as in the second example. $\sigma_2 >> \varepsilon_2\omega$, we get:

\begin{equation}\label{eq:final}
\begin{aligned}
    r = \sqrt{\frac{(\alpha_1-\alpha_2)^2+\beta_2^2}{(\alpha_1+\alpha_2)^2+\beta_2^2}} = \sqrt{\frac{\omega\varepsilon_1-\sqrt{2\varepsilon_1\omega\sigma_2}+\sigma_2}{\omega\varepsilon_1+\sqrt{2\varepsilon_1\omega\sigma_2}+\sigma_2}} \\
    \approx \sqrt{\frac{\sigma_2-\sqrt{2\varepsilon_1\omega\sigma_2}}{\sigma_2+\sqrt{2\varepsilon_1\omega\sigma_2}}},
\end{aligned}
\end{equation} 


When considering the high-loss assumption, the air's permittivity is quite low, therefore the final approximate expression on the right-hand side uses this fact to its advantage. The permittivity of air ($\varepsilon_1$), the angular frequency ($\omega$), and the electrical conductivity of the surface material ($\sigma_2$) are all affected by the high-loss version of the reflection coefficient, as indicated by Eq.\eqref{eq:final}. For high-loss cases, Eq.\eqref{eq:final} can be modified to derive surface electrical conductivity from radar reflection strength as shown below:

\begin{equation}\label{eq:final_final}
    \sigma_2 \approx \frac{2\varepsilon_1\omega(r^2+1)^2}{(r^2-1)^2}.
\end{equation}

Eq.~\eqref{eq:final_final}
indicates that the soil surface conductivity/resistivity ($\rho= 1/\sigma$) can be measured with the help of Synthetic Aperture Radar data.

\subsection{Dual Input Encoder Decoder UNet} \label{ref:implementation}
We design a architecture for the task at hand where the data has multiple modalities. Here the data modality corresponds to data from two different incidence angles from different flight lines. A trivial approach to the problem is to concatenate the data from multiple angles of incidence and pass them as one single input with multiple channels.
Another approach is to pass the channels of each incidence angle in separate input branches to the deep neural network and combine them at the decoder block. 

The intuition behind designing our architecture is that each input branch encoder can learn separate filters appropriate for the data of that input branch. Although each input branch has its own encoder pathway to the bottleneck, it also has skip connections to the common decoder block. We designed a separate encoder pathway for each incidence angle so that the input branch could take advantage of the data from different incidence angles. Our method is generalizable to data that has multiple modalities and the implementation of our method in this paper is specific to the case of data from two flight lines capturing multiple incidence data.
\subsubsection{Architecture}
The overall architecture (see Fig.~\ref{fig:DIUNet}) consists of two encoder pathways, one for each input branch and a common decoder block. Each encoder block consists of two successive 3x3 convolution layers with batch normalization followed by ReLU activation and a max pooling operation. As a result of max pooling after each encoder block, the feature size gets reduced by half until it reaches the bottleneck. Before performing the max-pooling operation, a skip connection copies the features to the decoder block from each input branch. 

No max pooling operations are performed once the bottleneck is reached and the decoder pathway starts. The decoder pathway consists of decoder blocks corresponding to each encoder block. The features from each encoder block through skip connections from each input branch are concatenated together in each decoder block after the bottleneck layer. 

The features are upsampled by transposed convolution that doubles the feature size when passed from one decoder block to the next. The output dimensions match the dimensions of the input, except for the number of channels. While the input dimensions for each input branch of DI-UNet is 128x128 with 23 channels, the output size is 128x128 with a single channel. The details of the channels and their significance are briefly mentioned in Table \ref{tab:polarimetric_parameter}. 

\section{Dataset \& Implementation details} \label{sec:datasetandimplementation}

\subsection{Dataset} \label{subsec:dataset}

\subsubsection{Area of study} \label{subsec:area}
\begin{figure}[htbp] 
\centerline{\includegraphics[width=80mm,scale=1]{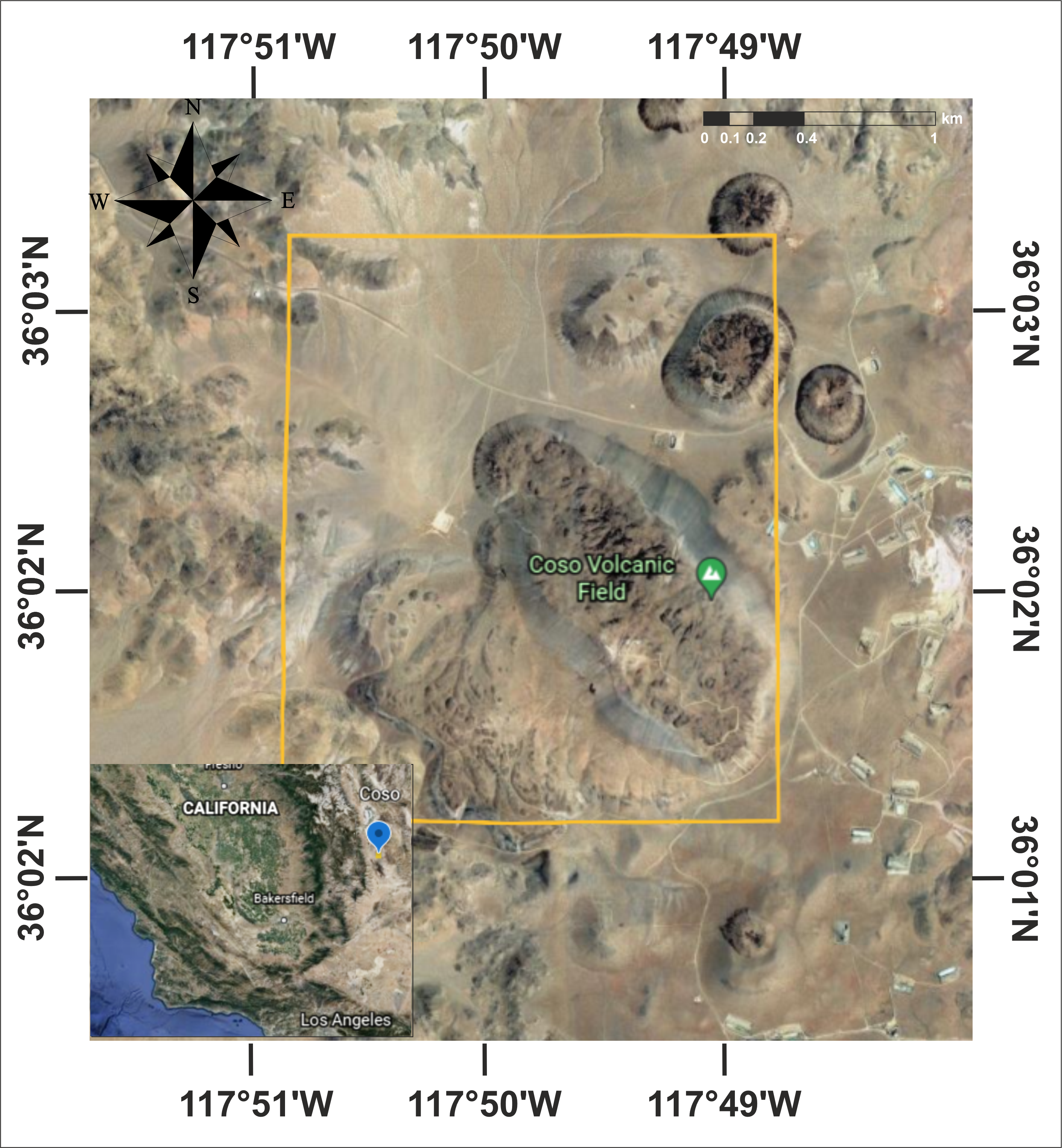}}
\caption{The study area considered, which belongs to coso geothermal area, California, USA. Within the orange rectangle is considered for the study.}
\label{fig:area}
\end{figure}

We have taken the Coso Geothermal Area, (CA, USA) as the primary observation area for our  pilot  study (Fig:~\ref{fig:area}). It is located in the central-eastern region of California within Inyo and Owen's counties. The geothermal energy available from this source has been used to generate electricity since 1987. 

\subsubsection{Remote sensing data}
Due to the all-weather availability, cloud penetration capabilities and sensitivity towards geometrical dielectric properties of the target, Synthetic Aperture Radar (SAR) data is gaining significant importance in geophysical parameter studies. 
In this context, the popular airborne sensor, Uninhabited Aerial Vehicle Synthetic Aperture Radar (UAVSAR) data is a great candidate because of its design to collect airborne repeat track polarimetric and interferometric SAR data. UAVSAR acquires L-band full polarimetric data at a center frequency of 1.26 GHz and 80 MHz bandwidth~\cite{rosen2006uavsar}. In this study, we have used the multi-look complex (MLC) ground range detected (GRD) product. The datasets were acquired on 10\textsuperscript{th} October 2019. We have considered data from two different flight lines (viz. \#32041 and \#14057) to leverage the advantage of multi-incidence angle data. 



\begin{figure}[!h]
\centering
    \subfloat[\label{} ]{          \includegraphics[width=0.49\columnwidth]{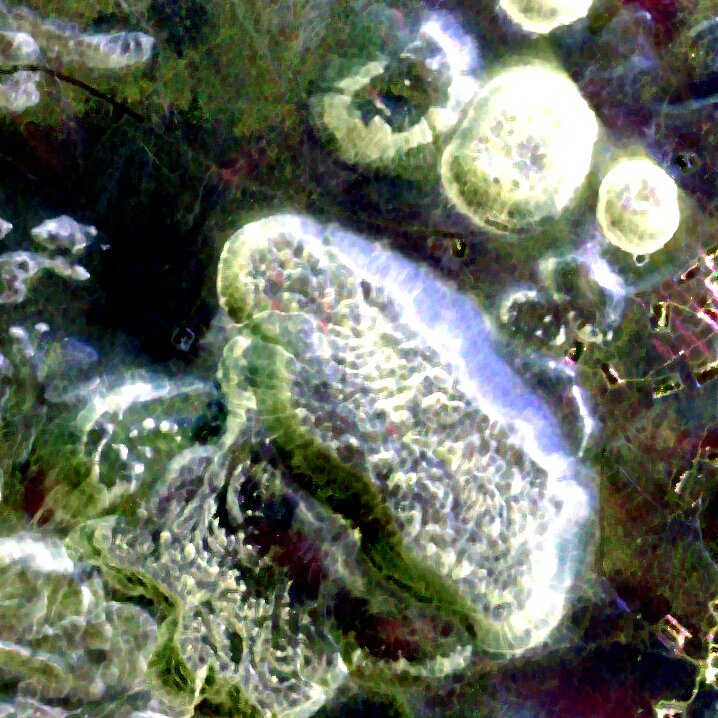}}
    \subfloat[\label{} ]{          \includegraphics[width=0.49\columnwidth]{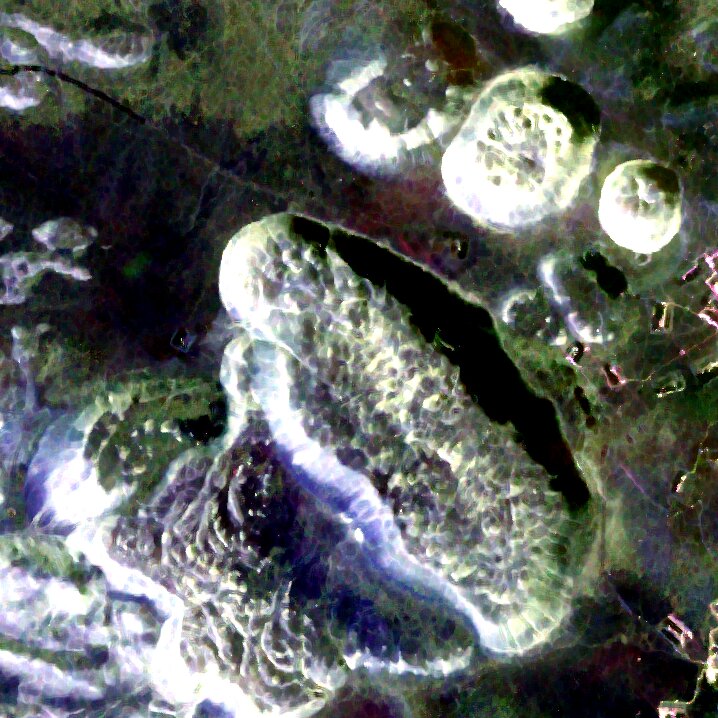}}	
    
	\caption{Study area UAVSAR: From left to right, PauliRGB generated from different flight line data (a) from flight line \#32041 (b) from flight line \#14057 }
	\label{fig:AU}
\end{figure}




\subsubsection{Magnetotelluric inversion resistivity data}

We have used publically available MT inversion data~\cite{blankenship2016west} of the Coso geothermal area. 125 MT stations and a single line of contiguous array profiling data were  acquired in the area, and inverted to a fully 3 dimensional resistivity model, where the observed impedence data were fit with the calculated data in a least square optimisation. Maxwell’s equations for 3D resistivity variations and plane wave source excitation in a discrete set of frequencies were used to solve and produce the computed model, which is the impedence tensor in field measurement~\cite{newman2008three}.
Each sample point was separated by 3 meters. The data was sliced at 450 feet along the z direction. We considered only the top surface slice for this study.
~\cite{blankenship2016west}. 


\subsection{Data preparation}
\subsubsection{UAVSAR}
The MLC data product is generated with a multilook factor of 3$\times$13 in range and azimuth direction respectively. The polarimetric coherency matrix $\langle [\textbf{T}] \rangle $ is extracted followed by a refined Lee filter with 5$\times$5 window size applied to despeckle the data. Furthermore, several polarimetric features are extracted using the PolSAR tools QGIS plugin~\cite{bhogapurapu2021polsar}. The details of the extracted features are provided in Table~\ref{tab:polarimetric_parameter}. Subsequently, the data covering the test site is divided into 70 equal patches with a size of 128 $\times$ 128 pixels.



\begin{table}[!ht]
\caption{Details of the polarimetric parameters/features used in this study.}
\label{tab:polarimetric_parameter}
\begin{adjustbox}{width=1\columnwidth}
\begin{tabular}{ll}
\toprule
\textbf{Short Form} & \textbf{Description} \\ \midrule
$\sigma^\circ_{\text{HH}},~\sigma^\circ_{\text{HV}},~\sigma^\circ_{\text{VV}}$ & Backscatter intensities \\
$\lambda_1,~\lambda_2,~\lambda_3$ & Eigen values \\

H, A, $\overline{\alpha}$ & Cloude decomposition parameters~\cite{cloude1997entropy} \\
$\beta,~\delta,~\gamma$ & Polarimetric intercorrelation parameters~\cite{cloude1997entropy} \\
$m_{\text{FP}}$ & Barakat degree of polarization~\cite{barakat1977degree} \\
$\theta_{\text{FP}}$ & Scattering type parameter based on degree of polarization ~\cite{dey2020target}\\
$P_s,~P_d,~P_v,~P_c$ & Model-free decomposition powers~\cite{dey2020target}\\
purity & scattering degree of purity~\cite{gil2007polarimetric}\\
depolarization & depolarization index~\cite{nafie2011vibrational}\\
Span & Total power\\
$H_I$, $H_P$ & Shannon entropy parameters\\
\bottomrule
\end{tabular}
\end{adjustbox}
\end{table}

\subsubsection{MT inversion data}
The resistivity data were sliced and converted from the CSV format to GeoTiff, using QGIS. The data was then superimposed with UAVSAR data and the entire area was split into 70 equal patches of size 128 $\times$ 128 pixels. Resistivity values in the surface range from 76 to 385 Ohm-m for the coso geothermal area.
\subsection{Implementation details}
\label{subsec:Implementation details}
The 23 extracted features from the UAVSAR are considered as the input to the network in 23 different bands, and resistivity band from the coso geothermal area as the ground truth (output). We used the L1 loss function between the prediction and the ground truth data as the objective function to optimize our model. To train our neural network model, we used Adam optimizer with an initial learning rate of 1$e^{-3}$ and a weight decay of 1$e^{-5}$. The beta coefficients used for computing running averages of gradient and its square are (0.9, 0.999). The eps of $1e^{-8}$ is added to the denominator to improve numerical stability. Due to data limitation, we used data augmentation techniques such as random cropping, vertical and horizontal flipping, reflected mode padding, and image resizing while training. For all the experiments involving models UNet, SegNet, PSPNet, and Dual-Input UNet, training was performed up to 200 epochs.
\section{Experiments and Evaluation}\label{sec:experimentsandevaluation}

\subsection{Evaluation and Error metrics}
\subsubsection{Evaluation metric}
We use metrics such as cosine similarity, explained variance, Pearson's correlation coefficient, and the coefficient of determination to evaluate our method. 
For these metrics larger values imply better performance.

\subsubsection{Error metrics}
There are error metrics like mean absolute error, mean absolute percentage error, root mean squared error and mean squared log error that can be used to evaluate the performance of methods. For error-based metrics, smaller values imply better performance.

\begin{figure*}[!ht]
\begin{minipage}[b]{0.23\textwidth}
\includegraphics[width=\textwidth]{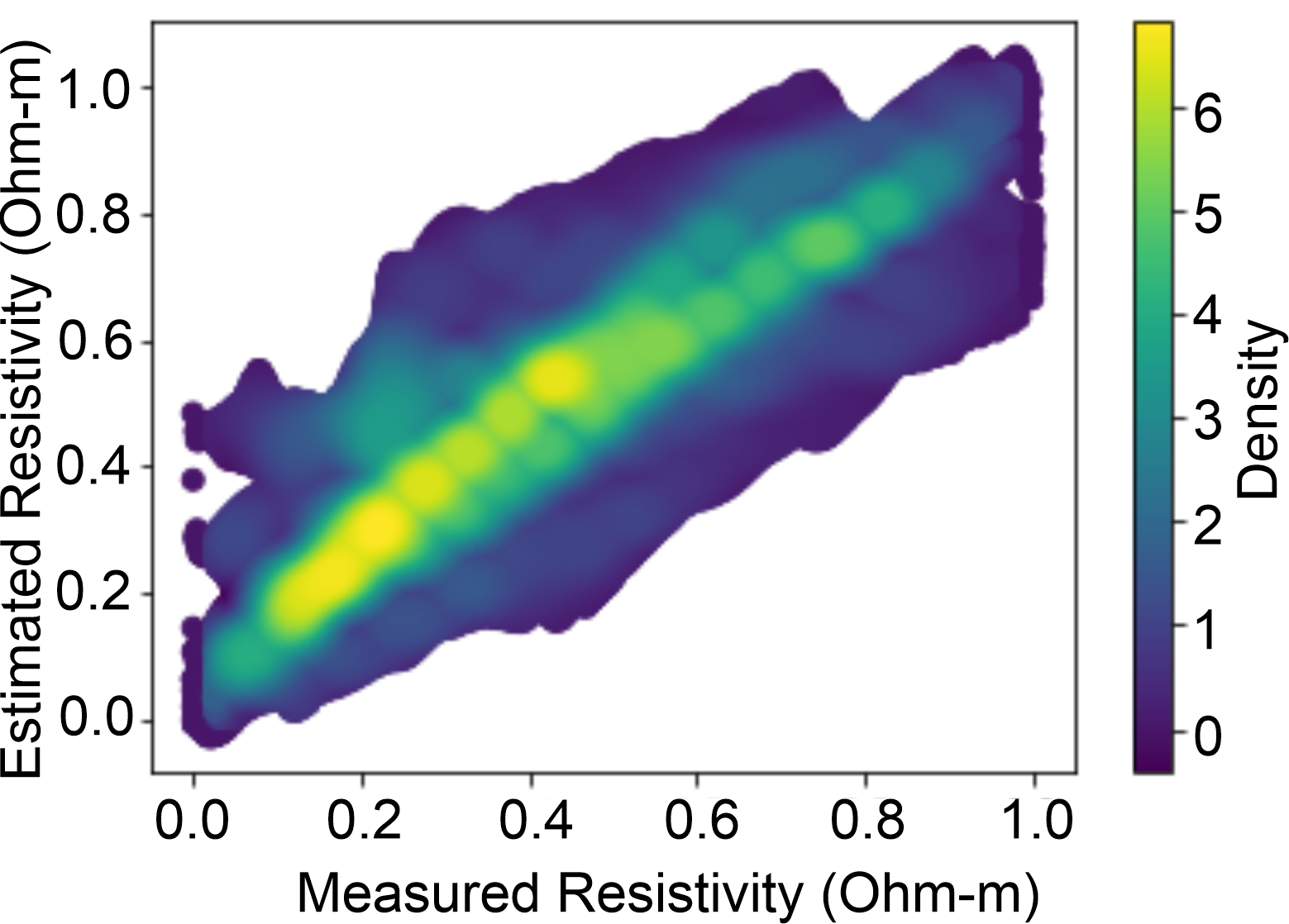}
\end{minipage}
\begin{minipage}[b]{0.23\textwidth}
\includegraphics[width=\textwidth]{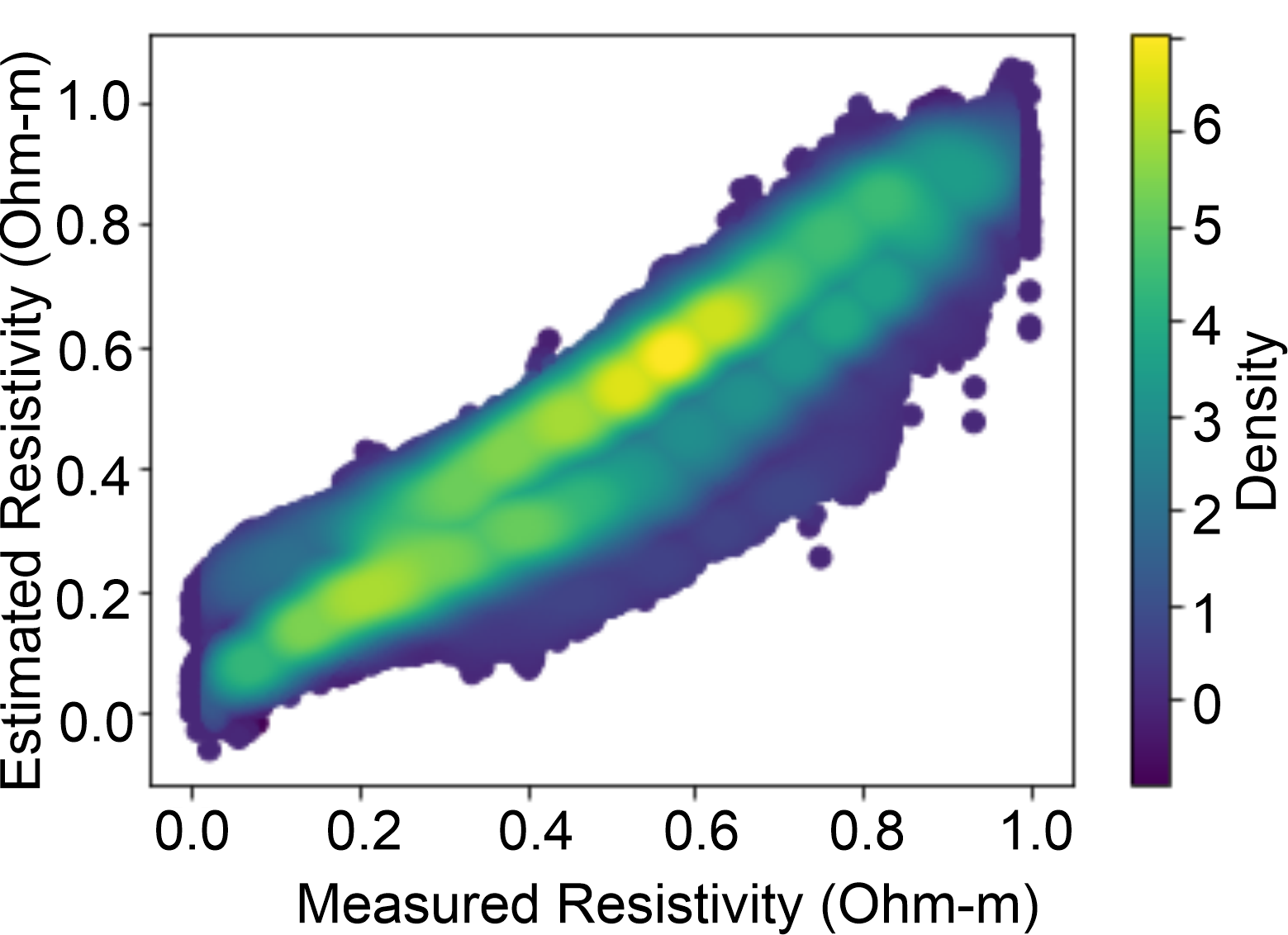}
\end{minipage}
\begin{minipage}[b]{0.23\textwidth}
\includegraphics[width=\textwidth]{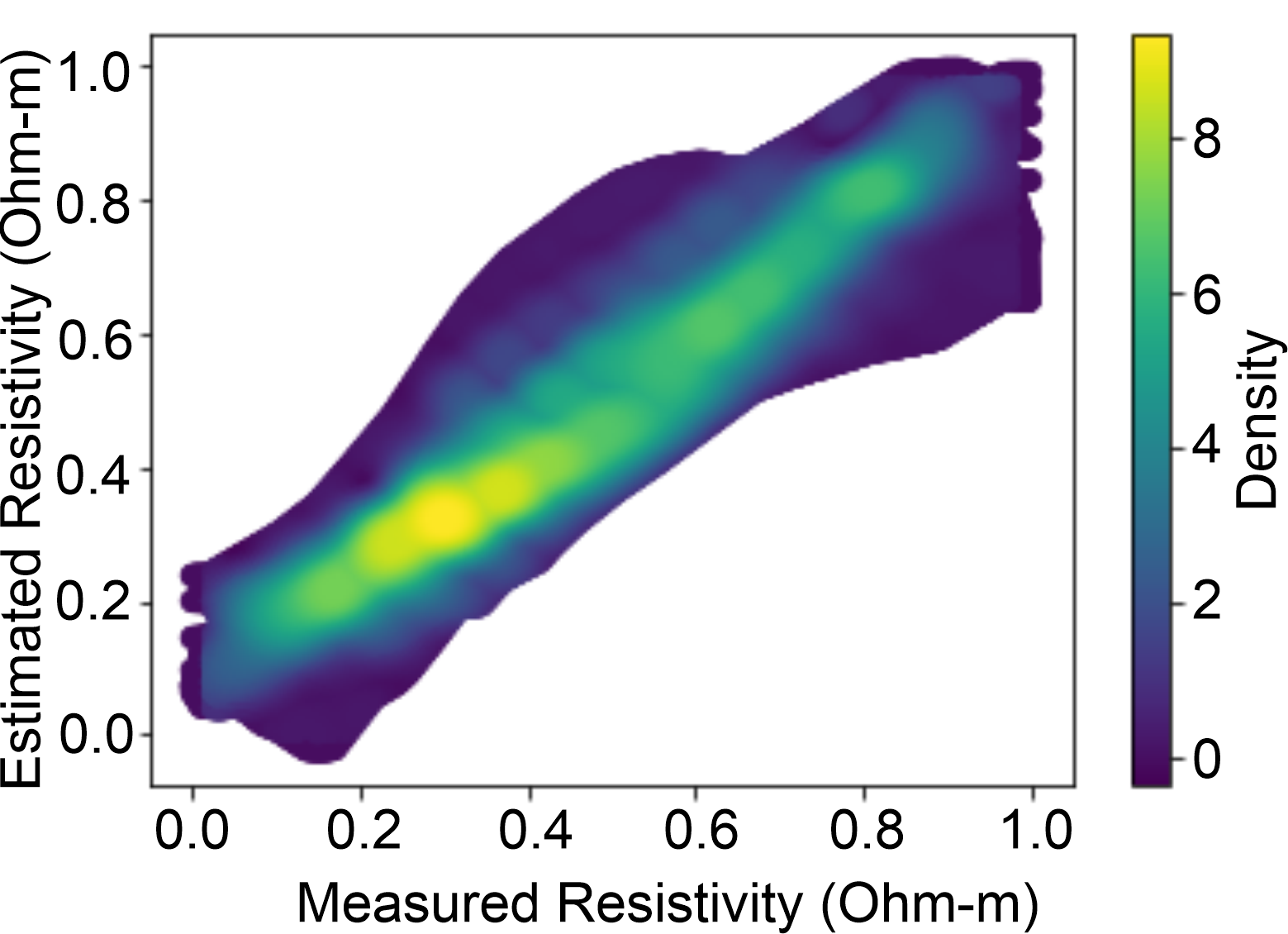}
\end{minipage}
\begin{minipage}[b]{0.23\textwidth}
\includegraphics[width=\textwidth]{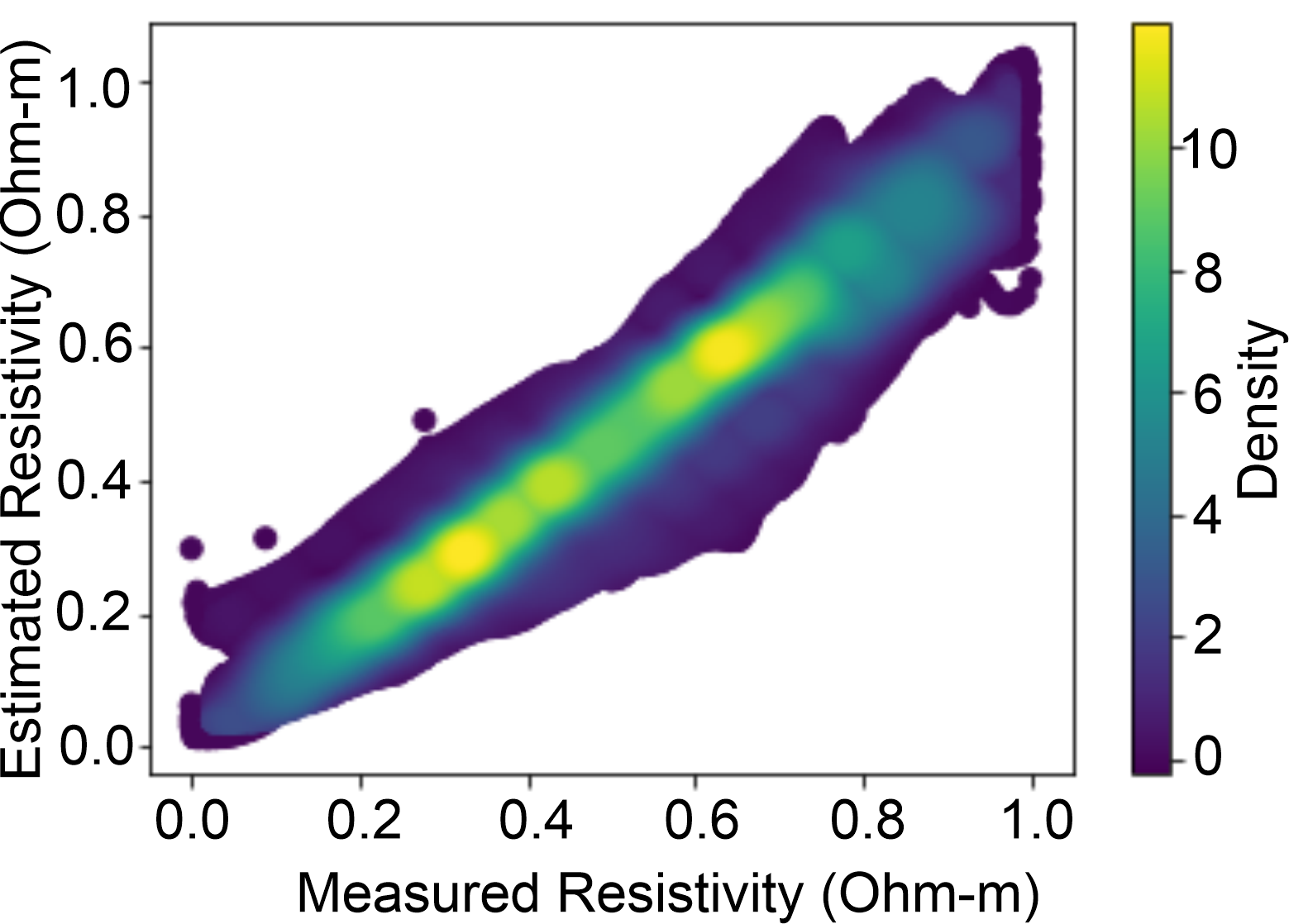}
\end{minipage}
\caption{Figures from left to right represents scatter plots of UNet, SegNet, PSPNet \& DI-UNet(Ours) respectively on test dataset}
\label{fig:scatter-plots}
\end{figure*}

\begin{table}[]
\caption{Comparison of UNet, SegNet and PSPNet on single incidence angle data (SAR1) using evaluation metrics(CS, EV, PCC, SCC and $R^2$)}
\label{tab:results-single-incidence-SAR1}
\centering
\begin{tabular}{l|ccccc}
\toprule
\textbf{DNN} &
\multicolumn{1}{l}{\textbf{CS}} &
\multicolumn{1}{l}{\textbf{EV}} &
\multicolumn{1}{l}{\textbf{PCC}} &
\multicolumn{1}{l}{\textbf{SCC}} &
\multicolumn{1}{l}{\textbf{$R^2$}} \\ \midrule
UNet & 0.9421 & 0.4114 & 0.6738 & 0.6714 & 0.4088 \\
SegNet & 0.9752 & \textbf{0.7500} & \textbf{0.8662} & \textbf{0.8872} & \textbf{0.7478} \\
PSPNet & \textbf{0.9774} & 0.7390 & 0.8630 & 0.8614 & 0.7349 \\ \bottomrule
\end{tabular}
\end{table}

\begin{table}[]
\caption{Comparison of UNet, SegNet, and PSPNet in single incidence angle data (SAR2) using evaluation metrics (CS, EV, PCC, SCC and $R^2$)}
\label{tab:results-single-incidence-SAR2}
\centering
\begin{tabular}{l|ccccc}
\toprule
\textbf{DNN} &
\multicolumn{1}{l}{\textbf{CS}} &
\multicolumn{1}{l}{\textbf{EV}} &
\multicolumn{1}{l}{\textbf{PCC}} &
\multicolumn{1}{l}{\textbf{SCC}} &
\multicolumn{1}{l}{\textbf{$R^2$}} \\ \midrule
UNet & 0.9428 & 0.4054 & 0.6694 & 0.6735 & 0.3932 \\
SegNet & 0.9687 & 0.6850 & 0.8296 & 0.8298 & 0.6761 \\
PSPNet & \textbf{0.9827} & \textbf{0.8098} & \textbf{0.9004} & \textbf{0.9020} &\textbf{ 0.8087} \\ \bottomrule
\end{tabular}
\end{table}


\begin{table}[]
\centering
\caption{Comparison of UNet, SegNet, and PSPNet on single incidence angle data (SAR1) using Errors metrics (MAE, MAPE, RMSE, MSLE)}
\label{tab:errors-single-incidence-SAR1}
\begin{tabular}{l|rrrr}
\toprule
\textbf{DNN} & \multicolumn{1}{l}{\textbf{MAE}} & \multicolumn{1}{l}{\textbf{MAPE}} & \multicolumn{1}{l}{\textbf{RMSE}} & \multicolumn{1}{l}{{\color[HTML]{000000} \textbf{MSLE}}} \\ 
\midrule
UNet & 0.1282 & 0.9122 & 0.0346 & 0.0155 \\ 
SegNet & 0.0818 & 0.5413 & \textbf{0.0133} & 0.0079 \\ 
PSPNet & \textbf{0.0749} & \textbf{0.4419} & 0.0137 & \textbf{0.0062} \\ 
\bottomrule
\end{tabular}

\end{table}

\begin{table}[]
\centering
\caption{Comparison of UNet, SegNet and PSPNet on single incidence angle data (SAR2) using error metrics (MAE, MAPE, RMSE, MSLE)}
\label{tab:errors-single-incidence-SAR2}
\begin{tabular}{l|rrrr}
\toprule
\textbf{DNN} & \multicolumn{1}{l}{\textbf{MAE}} & \multicolumn{1}{l}{\textbf{MAPE}} & \multicolumn{1}{l}{\textbf{RMSE}} & \multicolumn{1}{l}{{\color[HTML]{000000} \textbf{MSLE}}} \\ 
\midrule
SegNet  & 0.1013 & 0.5559 & 0.0174 & 0.0084 \\ 
UNet & 0.1420  & 0.7634 & 0.0367 & 0.0159 \\ 
PSPNet & \textbf{0.0747} & \textbf{0.7191} & \textbf{0.0106} & \textbf{0.0051} \\ 
\bottomrule
\end{tabular}

\end{table}

\begin{table}[]
\centering
\caption{Comparison of UNet, SegNet, PSPNet and DI-UNet on Multi Incidence Angle using evaluation metrics (CS, EV, PCC, SCC and $R^2$)}
\label{tab:results-full}

\begin{tabular}{l|ccccc}
\toprule
\textbf{DNN} &
\multicolumn{1}{l}{\textbf{CS}} &
\multicolumn{1}{l}{\textbf{EV}} &
\multicolumn{1}{l}{\textbf{PCC}} &
\multicolumn{1}{l}{\textbf{SCC}} &
\multicolumn{1}{l}{\textbf{$R^2$}} \\ \midrule
UNet & 0.9679 & 0.7054 & 0.8445 & 0.8446 & 0.6256 \\ 
SegNet & 0.9837 & 0.8286 & 0.9112 & 0.9114 & 0.8142 \\
PSPNet & 0.9878 & 0.8714 & 0.9335 & 0.9372 & 0.8661 \\
DI-UNet & \textbf{0.9927} & \textbf{0.9174} & \textbf{0.9578} & \textbf{0.9603} & \textbf{0.8714} \\ \bottomrule
\end{tabular}

\end{table}

\begin{table}[]
\centering
\caption{Comparison of UNet, SegNet, PSPNet and DI-UNet on Multi Incidence Angle using error metrics (MAE, MAPE, RMSE, MSLE)}
\label{tab:error-full}
\begin{tabular}{l|rrrr}
\toprule
\textbf{DNN} & \multicolumn{1}{l}{\textbf{MAE}} & \multicolumn{1}{l}{\textbf{MAPE}} & \multicolumn{1}{l}{\textbf{RMSE}} & \multicolumn{1}{l}{\textbf{MSLE}} \\ 
\midrule
UNet   & 0.1088 & 0.8866 & 0.0202                           & 0.0098 \\ 
SegNet & 0.0781 & 0.3434 & 0.0106                           & 0.0049 \\ 
PSPNet & 0.0640  & 0.5078 & \textbf{0.0069} & 0.0032 \\ 
DI-UNet & \textbf{0.0631} & \textbf{0.2843} & \textbf{0.0069} & \textbf{0.0029} \\ 
\bottomrule
\end{tabular}

\end{table}

\begin{figure}[!h]
\centering
    \subfloat{          \includegraphics[width=0.23\columnwidth]{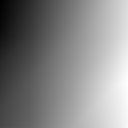}}
    \subfloat{          \includegraphics[width=0.23\columnwidth]{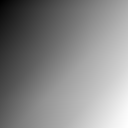}}
    \subfloat{          \includegraphics[width=0.23\columnwidth]{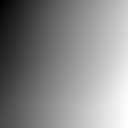}}
    \subfloat{          \includegraphics[width=0.23\columnwidth]{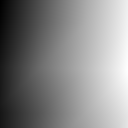}}

    \subfloat{          \includegraphics[width=0.23\columnwidth]{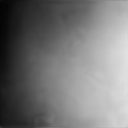}}
    \subfloat{          \includegraphics[width=0.23\columnwidth]{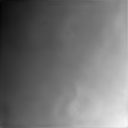}}
    \subfloat{          \includegraphics[width=0.23\columnwidth]{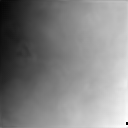}}
    \subfloat{          \includegraphics[width=0.23\columnwidth]{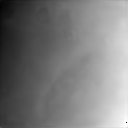}}
    
	\caption{Top row consists of the ground truth visualization, while the bottom row consists of the prediction visualizations corresponding to the ground truth images above. }
	\label{fig:GT_prediction}
\end{figure}

\subsection{Experiments:}

We conducted experiments on SAR data separately for each incidence angle using the methods UNet, SegNet and PSPNet. The results of the comparison using evaluation metrics cosine similarity, explained variance, Pearson's correlation coefficient, Spearman's correlation coefficient and $R^2$ are given in Table \ref{tab:results-single-incidence-SAR1} for the SAR1 data and Table \ref{tab:results-single-incidence-SAR2} for the SAR2 data. The results of comparison using metrics mean absolute error, mean absolute percentage error, root mean squared error and mean square log error are given in Table \ref{tab:errors-single-incidence-SAR1} for SAR1 data and Table \ref{tab:errors-single-incidence-SAR2} for SAR2 data.
We then combine the SAR1 and SAR2 data corresponding to the same region by concatenating them and passing them as input to the methods UNet, SegNet, and PSPNet, while our method DI-UNet takes the SAR1 and SAR2 data in different input branches. We conducted experiments to compare DI-UNet and other methods using the same metrics CS, EV, PCC, SCC, and R\textasciicircum{}2 and the results are given in Table \ref{tab:results-full}. We also conducted experiments to compare UNet, SegNet PSPNet and DI-UNet using error metrics MAE, MAPE, RMSE, and MSLE and the results are given in Table \ref{tab:error-full}. 
\section{Results and Analysis} \label{sec:results}

Figure \ref{fig:GT_prediction} visualizes the ground truth values and the corresponding estimations using our method. It is obvious from the qualitative comparison that the model predictions are nonsmooth, while the ground truth is smooth. This implies that our method has not overfitted to the underlying data and has done a fair job of modeling the relation between SAR data and the electrical resistivity.

The results from Figure \ref{fig:scatter-plots} indicate that our method has a single high density band corresponding to the diagonal to both axes, while other methods have multiple high density bands on either side of the diagonal. This indicates that the resistivity estimated by our method and the measured resistivity are very close, while other methods do not have an accurate prediction.

The scatter plots of estimated resistivity along the y axis is plotted against the measured resistivity given along x axis in Fig. \ref{fig:scatter-plots} for the different methods. The breadth of the density plot along the diagonal indicates the variance or divergence of the estimated resistivity from the measured resistivity. The larger the variance, greater the area covered under the scatter plot. In addition, the greenish or yellowish plot implies higher concentrations or higher density. If larger densities are located close to the diagonal, then the resistivity is estimated close to the measured resistivity. For an ideally performing method, the scatter plot should have maximum density of estimated resistivity to measured resistivity concentrated around the diagonal with a thin diagonal and small area of scatter plot.

We can observe from the scatter plots in Figure \ref{fig:scatter-plots} that DI-UNet records the largest density of correlation between measured and estimated resistivity ($>$10) among all the methods, while UNet, Segnet and PSPNet have a max density ranging between 6 and 9, where UNet has the least performnce while PSPNet has comparable performance to our method. We also observe that except for DI-UNet, all other methods have dual bands around the diagonal. This implies that the estimated resistivity of DI-UNet is aligned with the measured resitivity, while other methods have more variance in predicting the measured values. Among the compared methods UNet has the largest scatter plot area while our method has the least scatter plot area. This implies that our method has the least variance of predicted values with maximum correlation of estimated resistivity to measured resistivity. In other words, we can say that our method is able to model the underlying relationship between SAR data and measured resistivity.

Analyzing the quantitative results in Table \ref{tab:results-single-incidence-SAR1} and \ref{tab:results-single-incidence-SAR2} for metrics CS, EV, PCC, SCC, and R\textasciicircum{}2, we observe that SegNet performs better than UNet and PSPNet for SAR1 data, while PSPNet performs better than UNet and SegNet for SAR2 data. Similarly for error metrics MAE, MAPE, RMSE, MSLE in Table \ref{tab:errors-single-incidence-SAR1} and \ref{tab:errors-single-incidence-SAR2}, we observe that PSPNet performs better than UNet and SegNet for SAR1 data except in terms of RMSE, while PSPNet  performs consistently better than UNet and SegNet for SAR2 data. We can infer that it is sub-optimal to use the data from only one incidence angle, comparing the performance of UNet, SegNet and PSPNet trained only on SAR1 data (Table \ref{tab:results-single-incidence-SAR1} and \ref{tab:results-single-incidence-SAR2}) or SAR2 data (Table \ref{tab:errors-single-incidence-SAR1} and \ref{tab:errors-single-incidence-SAR2}) compared to Table \ref{tab:results-full} and Table \ref{tab:error-full}, trained on complete SAR data.

From Table \ref{tab:results-full} and Table \ref{tab:error-full}, we observe that our method has outperformed the existing methods in all evaluation and error metrics, with our closest competitor in performance being PSPNet. PSPNet uses a pyramid pooling module as a global contextual prior, while global average pooling is used as a global contextual prior in other segmentation methods. We attribute the performance of PSPNet due to use of pyramid pooling network. 

To conclude, methods SegNet and PSPNet are comparable to each other based on the evaluation metrics from Table \ref{tab:results-single-incidence-SAR1} for the SAR1 data and Table \ref{tab:results-single-incidence-SAR2} for the SAR2 data and the error metrics in Table \ref{tab:errors-single-incidence-SAR1} for SAR1 data and Table \ref{tab:errors-single-incidence-SAR2} for SAR2 data. However, based on the evaluation metrics from Table \ref{tab:results-full}, DI-Unet is consistently better than all other methods and based on the error metrics from Table \ref{tab:error-full}, DI-Unet is consistently better than all other methods except in RMSE metric where DI-Unet is equivalent to PSPNet.

\begin{table}[]
\centering
\caption{Ablation study for skip connections and input swapping using evaluation metrics (CS, EV, PCC, SCC, $R^2$)}
\label{tab:Ablation study}
\begin{tabular}{l|ccccc}
\toprule
\textbf{DNN variant} &
\multicolumn{1}{l}{\textbf{CS}} &
\multicolumn{1}{l}{\textbf{EV}} &
\multicolumn{1}{l}{ \textbf{PCC}} &
\multicolumn{1}{l}{\textbf{SCC}} &
\multicolumn{1}{l}{\textbf{R\textasciicircum{}2}} \\ \midrule
w/o SAR1 skip & 0.9797 & 0.7768 & 0.8877 & 0.8831 & 0.7767 \\ 
w/o SAR2 skip & 0.9822 & 0.8055 & 0.8982 & 0.8944 & 0.8022 \\ 
DI-UNet swap &
\multicolumn{1}{c}{0.9895} &
\multicolumn{1}{c}{0.8854} &
\multicolumn{1}{c}{0.9434} &
\multicolumn{1}{c}{0.9466} &
\multicolumn{1}{c}{0.8563} \\ 
DI-UNet & \textbf{0.9927} & \textbf{0.9174} & \textbf{0.9578} & \textbf{0.9603} & \textbf{0.8714} \\ \bottomrule
\end{tabular}

\end{table}

\begin{table}[]
\centering
\caption{Ablation study for skip connections and input swapping using error metrics (MAE, MAPE, RMSE, MSLE)}
\label{tab:Ablation study-error}
\begin{tabular}{l|cccc}
\toprule
\textbf{DNN variant} & \multicolumn{1}{l}{\textbf{MAE}} & \multicolumn{1}{l}{\textbf{MAPE}} & \multicolumn{1}{l}{\textbf{RMSE}} & \multicolumn{1}{l}{\textbf{MSLE}} \\ 
\midrule
w/o SAR2 skip & 0.0760 & 0.4765 & 0.0105 & 0.0049 \\ 
w/o SAR1 skip & 0.0793 & 0.4699 & 0.0118 & 0.0059 \\ 
DI-UNet swap & \multicolumn{1}{c}{0.0676} & \multicolumn{1}{c}{\textbf{0.2716}} & \multicolumn{1}{c}{0.0077} & \multicolumn{1}{c}{0.0036} \\ 
DI-UNet & \textbf{0.0631} & 0.2843 & \textbf{0.0069} & \textbf{0.0029} \\ 
\bottomrule
\end{tabular}

\end{table}

\subsection{Ablation Study}
We perform an ablation study to identify the critical components and the factors that contribute to the performance of our method. The baseline for all of our experiments is the Dual-Input UNet architecture with skip connections from both encoders to the decoder. While training we had fixed one input branch for single incidence angle data (SAR1) and the other branch for the other incidence angle data (SAR2). We validate whether this choice is relevant by randomly selecting either of the input branches and passing the incidence angle data in an input branch different from the last iteration. This makes sure that each input branch learns the filters to deal with the incidence angle data from both streams of data.
\\
We also consider another experiment to validate the importance of skip connections in each input branch. We consider two variants where there are no skip connections from one of the input branches to the decoder. In one variant, there are no skip connections from the SAR1 input branch, while in the other variant, there are no skip connections from the SAR2 input branch. The results of the comparison of the baseline with the three variants are given in Table \ref{tab:Ablation study} and Table \ref{tab:Ablation study-error}.

\section{Conclusions} \label{sec:conclusion}
This paper introduces a deep learning based model to obtain the surface resistivity/conductivity from the remote sensing SAR data. We studied the hypothesis on the Coso Geothermal area, which had both resistivity inversion data as well as aerial SAR data (UAVSAR). The preprocessed SAR data was considered as input to the network, and surface resistivity from the geothermal region was taken as ground truth. We then adapted UNet as a multi-input encoder-decoder architecture to accommodate the multi-pass and multi-look SAR data. Our proposed approach accomplished improved outcomes for the mapping of MT resistivity from SAR data. The qualitative and quantitative results support the superiority of our model architecture over other deep learning architectures. Future works can try obtaining more surface resistivity data, which can further improve the generalization capabilities of deep learning models for remote sensing derived surface resistivity.
\color{black}

\section*{Acknowledgment}
The authors are grateful to NASA Jet Propulsion Laboratory (JPL) for providing UAVSAR data. The authors would like to thank the ESA-STEP team and PolSARpro team for providing open-source software tools for SAR data processing. Bibin Wilson, Rajiv Kumar and Narayanarao Bhogapurapu  would like to acknowledge the support of MHRD, Govt. of India, towards their doctoral research. 

\bibliographystyle{IEEEtran}
\bibliography{ref}



%








\end{document}